\documentstyle[12pt]{article}
\topmargin - 1 cm
\begin{document}
\begin{center}
{ \large \bf PARTICLE PRODUCTION OF COHERENTLY OSCILLATING NONCLASSICAL INFLATON IN  FRW UNIVERSE }\\[.65cm]
{\small K. K. VENKATARATNAM  and   P. K. SURESH } \footnote { email: pkssp@uohyd.ernet.in \\
$ ^{*}$Corresponding address  :\\
 P.K.Suresh, School of Physics,University of Hyderabad,Hyderabad 500 046.India.} $^{*}$ \\
{ \it School of Physics, University of Hyderabad, Hyderabad-500 046.India.}\\[.4cm]
\end{center}

\begin{abstract}
We study
 particle production  of coherently oscillating inflaton  in the semiclassical
 theory of gravity by representing  inflaton in
coherent and squeezed state formalisms. A comparative study of the
inflaton in classical gravity with   coherent state inflaton in
semiclassical gravity is also presented.
\end{abstract}

\newpage
{ \small \section { Introduction}}
        The hot big bang model or standard cosmology is
spectacularly successful. In short, it provides reliable and
tested account of the history of the universe from about 0.01sec
after the big bang until today,  some 15 billion years later.
Despite its success, the hot big bang model left many features of
the universe unexplained. The most important of these are horizon
problem, singularity problem, flatness problem, homogeneity
problem, structure formation problem, monopole problem and so on.
All these problems are very difficult and defy solution within the
standard cosmology. Most of these problems have, in the past
couple of decades, been either completely resolved or considerably
relaxed in the context of one complete scenario, called
 inflationary scenario [1]. At present
there are  different versions [2-5] of the inflationary scenario.
The main feature of all these versions is known as the
inflationary paradigm. According to the simplest version of the
inflationary scenario, the universe in the past expanded almost
exponentially with time, while its energy density was dominated
by the effective potential energy density of a scalar field,
called the inflaton. Sooner or later, inflation terminated and
the inflaton field started quasiperiodic motion with slowly
decreasing amplitude. The universe was empty of particles after
inflation and particles of various kinds created due to the
quasiperiodic evolution  of the inflaton field. The universe
became hot again due the oscillations and decay of the created
particles of various kinds. Form on ,it can be described by the
hot big bang theory.

  Most of the inflationary scenarios are based on the classical gravity of the
Friedmann equation and the scalar field equation in the
Friedmann-Robertson-Walker (FRW) universe, assuming  its validity even at the
very early stage of the
universe. However, quantum effects of matter fields and quantum fluctuations
are expected to play a significant role in this regime, though quantum gravity
effects are still negligible. Therefore, the proper description of a
cosmological
model can be studied in terms of the semiclassical gravity of the semiclassical
Friedmann equation with quantized matter fields as the source of gravity.
Recently , the study of quantum properties of inflaton has been received
much attention in semiclassical theory of gravity and inflationary scenarios.
  [6,7].  In the new
inflation scenario [8] quantum effects of the inflaton were partially taken into
account by using one-loop effective potential and an initial thermal
condition.
In the stochastic inflation [9] scenario the inflaton was studied quantum
mechanically by dealing with the phase-space quantum distribution function and
the probability distribution [10]. The semiclassical quantum
gravity  seems to be a viable method throughout the whole non-equilibrium
quantum process from the pre-inflation period of hot plasma in thermal
equilibrium to the inflation period and finally to the matter-dominated period.

 The aforementioned studies show that results obtained in classical gravity are
quite different from those in semiclassical gravity. Though both
classical and quantum inflaton in the oscillatory phase of the
inflaton lead  the same power law expansion, the correction to
the expansion does not show any oscillatory behavior in
semiclassical gravity in contrast to the oscillatory behavior
seen in classical gravity. It is to be  noted that, the
coherently oscillating inflaton suffers from particle production.
Such studies reveal that quantum effects  and quantum phenomena
play an important role in inflation scenario and the related
issues. Recently, it has been found that nonclassical state
formalisms are quite useful to deal with quantum effects in
cosmology [11-21].

 The goal of the  present paper is to study a massive, minimal inflaton in the FRW universe
in the context of semiclassical gravity by representing the scalar field in
squeezed and coherent states. We study the
 particle production, of the oscillatory phase, of the inflaton in
coherent and squeezed state formalisms in the semiclassical
theory of gravity. We give a brief comparative study of the
inflaton in classical gravity with  the coherent state inflaton
in semiclassical gravity.

 { \small \section{Scalar field in FRW metric}}
 Consider a flat Friedmann-Robertson-Walker  spacetime with the line
 element   (with $c$=1)
 \begin{equation}
 ds^{2}=-dt^{2}+S^{2}(t) (dx^{2}+dy^{2}+dz^{2}),
 \end{equation}
where $S(t)$ is an unspecified positive function of $t$. One can consider,
the  universe with such a metric,  as an expanding
universe, although $S(t)$ need not be increasing with time.  The
metric is treated as an unquantized external
field.

 The equation governing the scalar field is the Klein-Gordon
equation [22]:
\begin{equation}
( g^{\mu\nu} \nabla_{\mu} \nabla_{\nu}-m^{2}) \phi=0\,\,,
\end{equation}
where  $g^{\mu\nu}$ is the inverse of the metric tensor $g_{\mu\nu}$,
$\nabla_{\mu}$ is the covariant derivative  and
$\mu$, $\nu$ = 0, 1, 2, 3\,.

Assume, minimal coupling between the gravity and the scalar field,
the massive scalar
field
 dynamics is governed by the Lagrangian density:
\begin{equation}
{\cal L} =-{1\over2}
\sqrt{(-g)}(g^{\mu\nu}\partial_{\mu}\phi\partial_{\nu}\phi+
m^{2} \phi^{2})\,\,,
\end{equation}
Using the metric (1), (3) becomes
\begin{equation}
   L={1\over2} S^{3}(t) \left({\dot{\phi}^{2}-S^{-2}(t)}
\sum_{j=1}^{3} (\partial_{j}\phi)^{2}-m^{2}\phi^{2}\right)
\end{equation}
and  (2) becomes
\begin{equation}
  \ddot\phi+3\frac{\dot{S}(t)}{S(t)}\dot\phi+m^{2}\phi -
\sum_{j=1}^{3}\partial_{j}^{2}\phi= 0,
\end{equation}
where overdot represents a derivative with respect to time and
$\partial_j \equiv \frac{
\partial}{\partial x_j}$ represents the spatial derivative.
The scalar field can be quantized [22] by defining momentum conjugate to $\phi$ as
\begin{equation}
\pi = \frac{\partial L}{\partial\dot{\phi}}
\end{equation}
and  following the canonical quantization procedure.
 The Hamiltonian of the scalar field can be  obtained by using
\begin{equation}
H= \pi\dot{\phi}- L \,\,.
\end{equation}
Therefore,  the Hamiltonian of the scalar field is :
\begin{equation}
H=\frac{\pi^{2}}{2S^{3}(t)}+{1\over2}S(t)\sum_{j=1}^{3}
(\partial_{j}\phi)^{2}+{1\over2}S^{3}(t)m^{2}\phi^{2}\,\,.
\end{equation}
In the present study, we consider only the homogeneous models of the
scalar field. For a homogeneous scalar field (inflaton), the spatial
derivatives of $\phi$ vanish, hence the Hamiltonian of the inflaton
can be written as,
\begin{equation}
H=\frac{\pi^{2}}{2S^{3}(t)}+{1\over2}S^{3}(t) m^{2}\phi^{2}\,\,.
\end{equation}
Therefore, the temporal component of the energy-momentum tensor for the
inflaton is obtained as
 \begin{equation}
T_{00}=S^{3}(t)({1\over2}\dot\phi^{2}+{1\over2} m^{2}\phi^{2})\,\,.
\end{equation}
We, next  discuss the physical and mathematical properties
of the coherent and squeezed state formalisms briefly.
{\section{Coherent states and squeezed states}}
Squeezed states and coherent states are important classes of quantum states,
well-known in quantum optics [23].
At a glance, it may appear  that coherent
and squeezed states formalisms and cosmology are two different branches of
physics having no connections. However, the mathematical and physical
properties of
these states find much use in the  study of  many issues in cosmology.
Recently, squeezed  and coherent states are being used as probes for
studying  the quantum effects in cosmology such as cosmological particle
creation [12], entropy generation [18], detection of
gravitational waves [24], inflationary scenario [17] etc. Coherent states
are considered  as
most classical, that can be generated from the vacuum state $\mid 0 \rangle$ by
the action of displacement operator. In the present study, we use single mode
coherent and squeezed states only. A single mode coherent state can be
defined [25] as
\begin{equation}
\mid\alpha\rangle=D(\alpha)\mid 0\rangle\,\,,
\end{equation}
where $D(\alpha )$ is the single mode displacement operator,  given by
\begin{equation}
D(\alpha)=\exp({\alpha a^{\dag}-\alpha^{*}a})\,\,.
\end{equation}
Here, $\alpha$ is a complex number and $a$, $a^{\dag}$ are respectively the
annihilation and creation operators, satisfying $[a,a^{\dag}]=1$. The action
of $a$ and $a^{\dag}$ on the
coherent state gives
\begin{eqnarray}
a\mid\alpha\rangle=&\alpha\mid\alpha\rangle\nonumber\\
a^{\dag}\mid\alpha\rangle=&\alpha^{*}\mid\alpha\rangle\,\,.
\end{eqnarray}
The single mode displacement operator given by (12) satisfy the following
properties.
\begin{eqnarray}
D^{\dag} a D=&a+\alpha \nonumber \\
D^{\dag} a^{\dag} D=&a^{\dag}+\alpha^{*}\,\,.
\end{eqnarray}
A squeezed state is generated by the action of the squeezing operator on any
coherent
state is also  on the vacuum state. Therefore, a single mode squeezed
state is defined [25] as
\begin{equation}
\mid \alpha,\xi \rangle = Z(r,\varphi) D(\alpha)|0 \rangle\,\,,
\end{equation}
with $Z(r,\varphi)$, the
single mode squeezing operator, given by,
\begin{equation}
Z(r,\varphi)=\exp{r\over2}(e^{-i\varphi} a^{2}-e^{i\varphi} a^{\dag 2})\,\,.
\end{equation}
Here, $r$ is the squeezing parameter, which
determines the strength of the squeezing and $\varphi$ is the squeezing
angle, which
determines the distribution between conjugate variables,
with  $0\leq r\leq \infty$ and $-\pi\leq\varphi\leq\pi$.
The  squeezing  operator satisfy the following properties
\begin{eqnarray}
Z^{\dag}a Z =& a \cosh r-a^{\dag}e^{i\varphi}\sinh r \\ \nonumber
Z^{\dag}a^{\dag} Z =&a^{\dag}\cosh r-a e^{-i\varphi}\sinh r\,\, .
\end{eqnarray}
By setting $\alpha$=0 in (15), one obtains the squeezed vacuum state,
and is defined  [25] as:
\begin{equation}
\mid\xi\rangle=Z(r,\varphi)\mid 0\rangle\,\,.
\end{equation}
The squeezed vacuum state is  a many-particle state and hence
the resulting field may be called classical. However, the statistical
properties of these states
greatly differ from the coherent states and therefore, this state is
considered as
highly non-classical  having no analog in classical physics.
   In the case of coherent states, the variance of the conjugate variables are
always equal to each other, while in a squeezed state one component of the noise
is always squeezed with respect to the other. Therefore, in  (x,p) plane, the
noise for the coherent state can be described by a circle and for the
 squeezed state as an ellipse.
{\section{Particle production of nonclassical inflaton in semiclassical gravity}}
 In the quantum theory of a matter field in curved space-time, the background
metric is usually treated as classical while the matter field is treated as
quantum. Two standard approaches towards the quantum field  have
been developed; one is the conventional field theoretical approach, the
other being  the canonical quantum gravity.  The correct theory of a quantum
fluctuating geometry and matter field  has not yet been  completely
developed; it would be meaningful to consider the semiclassical gravity theory
to study the quantum effect of matter field in a prescribed background metric.
The
semiclassical  approach is also useful to deal with problems in
cosmology, where quantum gravity effects are negligible. In the present study,
 we consider the oscillatory phase of inflaton after the inflation, where quantum
gravity effects are  negligible. Therefore the present
study  can be restricted
in the frame work of semiclassical  gravity.
 In semiclassical theory the Einstein equation can be written as
( here onwards we use the unit system $ c= \hbar $=1  and $ G=\frac{1}{m_{p}^2}$ ):
  \begin{equation}
G_{\mu\nu}=\frac{8\pi}{m^{2}_{p}} \langle {\hat{T}}_{\mu\nu}\rangle\,\,.
\end{equation}
The source of the gravitational field is the
 quantum , a scalar field $\phi$,
governed by the time-dependent Schr$\ddot{o}$dinger equation,
\begin{equation}
i\frac{\partial}{\partial t} \Phi(\phi,t)=\hat{H}_{m}(\phi,t)
\Phi(\phi,t)\,\,.
\end{equation}
Here, $\hat{H}_m$ represents the Hamiltonian for the matter field.

 As mentioned earlier, we consider a massive inflaton, minimally coupled to a spatially flat
FRW universe with the metric (1). Therefore, the
time-time component of the classical gravity is now the classical Einstein
(or Friedmann) equation
\begin{equation}
\left(\frac{\dot{S}}{S}\right)^2=\frac{8\pi}{3m^{2}_{p}} \frac{T_{00}}{S^3(t)}
\,\,,
\end{equation}
where $T_{00}$ is the energy density of the inflaton, given by (10).
The classical equation of motion for the inflaton is obtained from (5):
\begin{equation}
  \ddot{\phi}+3\frac{\dot{S}(t)}{S(t)}\dot\phi+m^{2}\phi= 0\,\,.
\end{equation}
In  the cosmological context, the classical Einstein equation (21) means that
the
Hubble constant, $H=\frac{\dot{S}}{S}$, is determined by the energy density of
the dynamically evolving inflaton as described by (22).
In the semiclassical theory, the Friedmann equation can be written as:
\begin{equation}
\left(\frac{\dot{S}}{S}\right)^2=\frac{8\pi}{3m^{2}_{p}} \frac{1}{S^3(t)}
\langle \hat{H}_m \rangle\,\,,
\end{equation}
where $\langle \hat{H}_m \rangle$ represent the expectation value of the Hamiltonian
of the scalar field in a quantum state under consideration.

Consider, a massive inflaton, minimally coupled to
 the spatially flat FRW metric .
 The  inflaton  can be
described by the time dependent harmonic oscillator, with the
Hamiltonian given in (9).
To study,  the semiclassical Friedmann equation, the
expectation value
the  Hamiltonian (9) to be computed, in a quantum state under consideration. Therefore (9) becomes:
\begin{equation}
\langle\hat{H}_{m} \rangle=\frac{1}{2S^{3}}\langle\hat{\pi}^2\rangle+
\frac{m^{2}S^{3}}{2}\langle\hat{\phi}^2\rangle\,\, .
\end{equation}
The eigenstates of the  Hamiltonian are the Fock states:
\begin{equation}
\hat{a^{\dag}}(t)\hat{a}(t)|n,\phi,t \rangle = n|n,\phi,t \rangle\,\,,
\end{equation}
where $a^{\dag}$,$a$  are the creation and annihilation operators
obeying boson commutation relations $[a,a^{\dag}]=1$, the other
combinations being zero. These can respectively be written as:
\begin{eqnarray}
\hat{a}(t)&=&\phi^*(t) \hat{\pi} - S^{3}\dot{\phi}^*(t)
\hat{\phi},\nonumber\\
\hat{a^{\dag}}(t)&=&\phi(t) \hat{\pi}-S^{3} \dot{\phi}(t)
\hat{\phi}\,\,.
\end{eqnarray}
As an  alternative to the $n$  representation of the inflaton,  next consider
the inflaton  in the coherent  state formalism,
therefore the semiclassical Einstein equation can be expressed in terms of the
coherent state parameter.

 From  (11), (14) and (26), we get
\begin{equation}
\langle\hat{\pi}^{2}\rangle_{cs} =S^{6}\left[2(|\alpha|^{2}+{1\over2})
\dot{\phi}^*\dot{\phi}-\alpha^{*2}\dot{\phi}^{*2}
-\alpha^{2}\dot{\phi}^{2}\right]
\end{equation}
  and
\begin{equation}
\langle\hat{\phi}^{2}\rangle_{cs}=2\left(|\alpha|^{2}+ {1\over2}\right)
\phi^*\phi-\alpha^{*2}
\phi^{*2}-\alpha^{2}\phi^{2}\,\,.
\end{equation}
 Substituting   (27) and (28) in (24),
the expectation value of the Hamiltonian  obtained for the coherent state as:
\begin{equation}
\langle \hat{H_{m}}\rangle_{cs}=S^{3}\left[(|\alpha|^{2}+{1\over2})(\dot{\phi}^*
\dot{\phi}+m^{2}\phi^*\phi)-{1\over2}\alpha^{*2}[\dot{\phi}^{*2}+m^{2}
\phi^{*2}]
-{1\over2}\alpha^{2}(\dot{\phi}^{2}+m^{2}\phi^{2})\right]\,\,.
\end{equation}
In  (29),  $\phi$ and $\phi^*$ satisfy (22) and the
Wronskian condition, given by:
\begin{equation}
S^3(t)\left(\dot{\phi^*}(t)\phi(t)-\phi^{*}(t) \dot{\phi}(t) \right)=i\,\,.
\end{equation}
The Wronskian and the boundary conditions, fix the normalization constants
of the  two independent solutions.

Transform the solution
in the following form
\begin{equation}
\phi(t)=\frac{1}{S^{3\over2}}\psi(t),
\end{equation}
thereby obtaining
\begin{equation}
\ddot{\psi}(t)+\left(m^{2}-{3\over4}\left(\frac{\dot{S}(t)}{S(t)}\right)^{2}
-{3\over2}
\frac{\ddot{S}(t)}{S(t)}\right)\psi(t)=0\,\,.
\end{equation}
Next, focus on the oscillatory phase of the inflaton after inflation. In the
parameter region satisfying  the inequality
\begin{equation}
m^2 > \frac{3\dot{S}^2}{4S^2}+\frac{3\ddot{S}}{2S},
\end{equation}
the inflaton has an oscillatory solution of the form
\begin{equation}
\psi(t)=\frac{1}{\sqrt{2w(t)}}\exp(-i\int w(t)dt)\,\,.
\end{equation}
With
\begin{equation}
w^{2}(t)=m^{2}-{3\over4}\left(\frac{\dot{S}(t)}{S(t)}\right)^{2}-{3\over2}
\frac{\ddot{S}(t)}{S(t)}+{3\over4}\left(\frac{\dot{w}(t)}{w(t)}\right)^{2}-
{1\over2}\frac{\ddot{w}(t)}{w(t)}\,\,.
\end{equation}

Next, consider the particle production of the inflaton, in
coherent  and squeezed states formalisms, in semiclassical theory
of gravity. First,  consider the Fock space which has a one
parameter dependence on the cosmological time $t$. The number of
particles at a later time $t$ produced from the vacuum at the
initial time $t_{0}$ is given by
\begin{equation}
N_0(t,t_0)=\langle 0,\phi,t_0\mid\hat{N}(t)\mid 0,\phi, t_0\rangle  ,
\end{equation}
Here, $\hat{N}(t)=a^{\dag}a$ and its expectation value can
be calculated by using (26) as
\begin{equation}
\langle \hat{N}(t) \rangle = S^6 \dot{\phi} \dot{\phi^*} \langle \hat{\phi^2}  \rangle
+ \phi \phi^* \langle \hat{\pi}^2   \rangle - S^3 \phi
\dot{\phi}^* \langle \hat{\pi} \hat{\phi} \rangle-S^3 \dot{\phi} \phi^* \langle
\hat{\phi} \hat{\pi}
\rangle  \,\,.
\end{equation}
 $\langle \hat{\pi}^2 \rangle$ , $\langle \hat{\pi} \hat{\phi} \rangle$
, $\langle \hat{\phi}\hat{\pi}\rangle$ and $\langle\hat{\phi}^2\rangle$ are
respectively obtained  as:
\begin{eqnarray}
\langle\hat{\pi}^2\rangle&=&S^6\dot{\phi}^*\dot{\phi},  \\ \nonumber
\langle\hat{\pi}\hat{\phi}\rangle&=&S^3 \dot{\phi} \phi^*, \\ \nonumber
\langle\hat{\phi}\hat{\pi}\rangle&=&S^3\phi\dot{\phi}^*,  \\ \nonumber
\langle\hat{\phi}^2\rangle&=&\phi^* \phi\,\,.
\end{eqnarray}
  Therefore, substituting (38),  in (37),  we get
\begin{equation}
N_0(t,t_0)=S^{6}|\phi(t)\dot{\phi}(t_0)-\dot{\phi}(t)\phi(t_0)|^{2}\,\,.
\end{equation}
 Using the  following approximation
ansatzs
\begin{equation}
w_{0}(t)=m
\end{equation}
and
\begin{equation}
S_{0}(t)=S_{0}t^{2\over3}\,\, ,
\end{equation}
and (31),
the number of particles
created at a later time $t$  from the vacuum state at the initial time
 $t_0$ in the limit $mt_0$,  $mt>1$ can be computed and is [7] given by:
\begin{eqnarray}
N_0(t,t_0)&=&\frac{1}{4 w(t) w(t_0)}\left(\frac{S(t)}{S(t_0)}\right)^3
\left[\frac{1}{4}\left(3\frac{\dot{S}(t)}{S(t)}-3\frac{\dot{S}(t_0)}{S(t_0)}
-\frac{\dot{w}(t)}{w(t)}+\frac{\dot{w}(t_0)}{w(t_0)}\right)^2\right.\nonumber\\
&&+\left.(w(t)-w(t_0))^2
\right]\nonumber\\
& \simeq &  \frac{ {( t-t_{0})}^2 } { 4 m^2 t_{0}^4} \, .
\end{eqnarray}
The
 expectation values of $\langle \hat{\pi}^2 \rangle$ , $\langle
\hat{\pi} \hat{\phi} \rangle$
, $\langle \hat{\phi}\hat{\pi}\rangle$  and  $\langle\hat{\phi}^2\rangle$
 in the coherent state are obtained as:
 \begin{eqnarray}
\langle\hat{\pi}^2\rangle_{cs}&=&S^6\left[
\left(2|\alpha|^2+1\right)
\dot{\phi}^*(t_0)
\dot{\phi}(t_0)-\alpha^{*2} \dot{\phi}^{*2}(t_0)-\alpha^2\dot{\phi}^2(t_0)
\right]\,\,,
\\ \nonumber
\langle\hat{\phi}^2\rangle_{cs}&=&\left(2|\alpha|^2+1\right)
\phi^*(t_0)\phi(t_0)-
\alpha^{*2} \phi^{*2}(t_0)-\alpha^2 \phi^2(t_0)\,\,,
\\ \nonumber
\langle\hat{\pi}\hat{\phi}\rangle_{cs}&=&S^3\left[|\alpha|^2\dot{\phi}^*(t_0)\phi(t_0) +\left(|\alpha|^2 +1 \right)
 \dot{\phi}(t_0)\phi^*(t_0)-\alpha^2\dot{\phi}(t_0)\phi(t_0) -\alpha^2 \dot{\phi}^*(t_0)
 \phi(t_0)\right],
\\ \nonumber
\langle\hat{\phi}\hat{\pi}\rangle_{cs}&=&S^3\left[
|\alpha|^2\phi^*(t_0)\dot{\phi}(t_0) +\left(|\alpha|^2+1\right)
 \phi(t_0)\dot{\phi}^*(t_0)-\alpha^2\phi(t_0) \dot{\phi}(t_0) -\alpha^{*2}\phi(t_0)
\dot{\phi}^*(t_0)\right]\,\,.
\end{eqnarray}
 Substituting (43) in (37) and using (40) and (41),
the number of particles at a later time $t$ produced from the
coherent state at the initial time $t_0$ is obtained as:
\begin{equation}
N_{cs}=(2|\alpha|^2+1) N_0(t,t_0)+|\alpha|^2-S^{6}
\alpha^{*2}E-S^6 \alpha^{2}F\,\,.
\end{equation}
Here, $N_0(t,t_0)$ is given by  (42) and
 \begin{eqnarray}
E&=&\phi(t)\phi^*(t) \dot{\phi^{*2}}(to)-\phi(t)\dot{\phi^*}(t)\dot{\phi^*}
(t_0)\phi(t_0)-\dot{\phi}(t)\phi^*(t)\phi(t_0)\dot{\phi^*}(t_0)+\dot{\phi}(t)
\dot{\phi^*}(t)\phi^{*2}(t_0) \\ \nonumber
&=&\frac{1}{4 w(t)w(t_0) S^3(t) S^3(t_0)}\left[\exp(2i\int w(t_0) dt_0
\left( {\cal{A}}_{2}-w^2(t_0)-iw(t_0)\right.\right. \\ \nonumber
&&\times\left.\left.
\left(\frac{3\dot{S}(t_0)}{S(t_0)}+\frac{\dot{w}(t_0)}{w(t_0)}\right)
+{\cal{A}}_{1}+w^2(t) \right)
-2 {\cal{C}}_{2}\left[\frac{3\dot{S}(t)}{S(t)} +\frac{\dot{w}(t)}{w(t)} \right]
\right] ,
\end{eqnarray}
\begin{eqnarray}
F&=&\phi(t)\phi^*(t) \dot{\phi^2}(to)-\phi(t)\dot{\phi^*}(t)\dot{\phi}
(t_0)\phi(t_0)-\dot{\phi}(t)\phi^*(t)\phi(t_0)\dot{\phi}(t_0)+\dot{\phi}(t)
\dot{\phi}^*(t)\phi^2(t_0)\,\,.\\ \nonumber
&=&\frac{\exp(-2i\int w(t_0) dt_0}{4 w(t) w(t_0) S^3(t) S^3(t_0)}\left[
\left[{\cal{A}}_{2}-w^2(t_0)+iw(t_0)
\left(\frac{3\dot{S}(t_0)}{S(t_0)}+\frac{\dot{w}(t_0)}{w(t_0)}\right)
\right]\right. \\  \nonumber
&&-\left[\frac{3\dot{S}(t_0)}{S(t_0)} +\frac{\dot{w}(t_0)}{w(t_0)}+
2iw(t_0) \right] {\cal{C}}_{1} \\ \nonumber
&&-\left.\left[\frac{3\dot{S}(t)}{S(t)} +\frac{\dot{w}(t)}
{w(t)}+2iw(t)\right] {\cal{C} }_{2}
+{\cal{A}}_{1}+w^2(t)\right].
\end{eqnarray}

where,
\begin{equation}
{\cal{A}}_{1}=\frac{1}{4}\left[3\frac{\dot{S}(t)}{S(t)}+\frac{\dot{w}(t)}
{w(t)}\right]^2,
\end{equation}
\begin{equation}
{\cal{A}}_{2}=\frac{1}{4}\left[3\frac{\dot{S}(t_0)}{S(t_0)}+\frac{\dot{w}(t_0)}
{w(t_0)}\right]^2,
\end{equation}
\begin{equation}
{\cal{C}}_{1}=\frac{3}{4} \frac{\dot{S}(t)}{S(t)}+\frac{1}{4} \frac{\dot{w}(t)}
{w(t)}-\frac{1}{2} iw(t)\,\,,
\end{equation}
and
\begin{equation}
{\cal{C}}_{2}=\frac{3}{4} \frac{\dot{S}(t_0)}{S(t_0)}+\frac{1}{4} \frac{\dot{w}(t_0)}
{w(t_0)}-\frac{1}{2} iw(t_0).
\end{equation}

 Substituting (45) and (46) in (44), then
using approximation ansatzs,  the
expression  becomes:
\begin{eqnarray}
N_{cs}=&&(2|\alpha|^2+1) N_0(t,t_0)+|\alpha|^2-\left(
\frac{\alpha^2}{4m^2}\right)
\left(\frac{t}{t_0}\right)^2\nonumber\\
&&\times\left[\exp(2i\int(\theta-mt_0)\left(\frac{1}{t^2_0}-3m^2
+\frac{2im}{t_0}-\frac{2}
{tt_0}+im\frac{(t-t_0)}{tt_0}\right) \right. \\ \nonumber
&&+\left.\exp(-2i\int(\theta-mt_0)\left(\frac{1}{t^2_0}-\frac{2im}{t_0}
+\frac{1}{t^2}\right)
-\exp(-2i\theta)\left(\frac{2}{tt_0}-\frac{2im}{t}\right)\right]\,\,.
\end{eqnarray}
Therefore, the number of particles at a later time $t$ produced from the
coherent state at the initial time $t_0$, in the limit $mt_0$, $mt>1$, is
obtained by setting $\theta=mt_0=0$ and dropping the imaginary terms:
\begin{eqnarray}
N_{cs} \simeq &&(2|\alpha|^2+1) N_0(t,t_0)+|\alpha|^2-\left(
\frac{\alpha^2}{4m^2}\right)
\left(\frac{t}{t_0}\right)^2
\left[\frac{2}{t^2_0}+\frac{1}{t^2}-3m^2
-\frac{4}{tt_0}\right]\,\,.
\end{eqnarray}
 Substituting $N_0(t,t_0)$ from (42) in (52),we get
\begin{eqnarray}
N_{cs} \simeq &&(2|\alpha|^2+1)\frac{(t-t_0)^2}
{4m^2t^4_0}+|\alpha|^2-\left(
\frac{\alpha^2}{4m^2 t^4_0}\right)
\left[2t^2+t^2_0-3t^2_0t^2m^2-4tt_0\right]\,\,.
\end{eqnarray}
Analogously, the number of particles at a later time $t$ produced  from
 squeezed vacuum
state at the initial time
$t_0$ can be obtained by using the following equations,
 \begin{eqnarray}
\langle\hat{\pi}^2\rangle_{svs}& = &  S^6\left[
\left(2\sinh^2r+1\right)
\dot{\phi}^*(t_0)
\dot{\phi}(t_0)+  \sinh r \cosh r ( e^{-i\varphi}
\dot{\phi}^{*2}(t_0)  + e^{i\varphi} \dot{\phi}^2(t_0) )
\right],  \nonumber \\
\langle\hat{\phi}^2\rangle_{svs}&=&\left(2\sinh^2r+1\right)
\phi^*(t_0)\phi(t_0)+\sinh r\cosh r (
e^{-i\varphi} \phi^{*2}(t_0)+e^{i\varphi}
\phi^2(t_0)), \\ \nonumber
\langle\hat{\pi}\hat{\phi}\rangle_{svs}&= & S^3\left[
\sinh^2r\dot{\phi}^*(t_0)\phi(t_0) +\cosh^2r
 \dot{\phi}(t_0)\phi^*(t_0)  +  \right. \\ \nonumber
&& \left.  \sinh r\cosh r (
e^{-i\varphi}
 \dot{\phi}^*(t_0) \phi(t_0)
+ e^{i\varphi} \dot{\phi}(t_0)\phi(t_0) )
 \right],\\
\langle\hat{\phi}\hat{\pi}\rangle_{svs}& = & S^3\left[
 \sinh^2r\phi^*(t_0)\dot{\phi}(t_0) +\cosh^2r
 \phi(t_0)\dot{\phi}^*(t_0) \right.  \nonumber  \\
& &+\left.\sinh r\cosh r  (
e^{-i\varphi}
\phi(t_0) \dot{\phi}^*(t_0)
+ e^{i\varphi} \phi(t_0) \dot{\phi}(t_0) )
 \right], \nonumber
\end{eqnarray}
   and (54) in (37). The
  number of particles produced in squeezed  vacuum
state  is then  obtained as:
\begin{equation}
N_{svs}=(2\sinh^2 r+1) N_0 (t,t_0)+\sinh^2 r +S^6 \sinh r \cosh r (
e^{-i\varphi}
 E+ e^{i\varphi}  F )\,\,,
\end{equation}
where $N_0(t,t_0)$ is given by (42), $E$ and $F$ are respectively given by (45) and (46) .

By setting $\varphi =2mt_{0} = 0 $ and dropping the imaginary terms , the  above equations becomes:
\begin{eqnarray}
N_{svs} \simeq (2\sinh^2 r+1)\frac{(t-t_0)^2}
{4m^2t^4_0}  +\sinh^2 r
+\frac{\sinh r \cosh r}{4m^2 t^4_0}\left[2t^2+t^2_0-4tt_0-3m^2t^2_0t^2
-2t_0t^2\right]\,\,.
\end{eqnarray}
Similarly
the number of particles at a later time $t$ produced from the
squeezed state at the initial time $t_0$ is obtained as:
\begin{eqnarray}
N_{ss}&=&(2\sinh^2 r + 1+2|\alpha|^2) N_0 (t,t_0)
+(\sinh^2 r+|\alpha|^2)   \nonumber  \\
&&+ S^6  \sinh r \cosh r ( e^{-i\varphi} E
+ e^{i\varphi}  F )
-S^{6}\alpha^{*2} E-S^6\alpha^2 F\,\,, \\ \nonumber
&\simeq &(2\sinh^2 r+2|\alpha|^2+1)\frac{(t-t_0)^2}
{4m^2t^4_0}  +\sinh^2 r+|\alpha|^2 \\ \nonumber
&&+\frac{\sinh r \cosh r}{4m^2 t^4_0}\left[2t^2+t^2_0-4tt_0-3m^2t^2_0t^2
-2t_0t^2\right] \\ \nonumber
&&-\left(\frac{\alpha^2}{4m^2 t^4_0}\right)
\left[2t^2+t^2_0-4tt_0-3m^2t^2_0t^2\right]\,\,.
\end{eqnarray}
   The above expression, when  $r=0$,  equals the number of
particles produced in the coherent state,
and for  $\alpha=0$, it equals  the number of particles
produced
in the squeezed vacuum state.
{\section{Comparison with classical gravity}}
Here, we briefly discuss some aspects of the inflaton in classical
gravity with semiclassical gravity, in its oscillatory regime.
Since coherent states are considered as
the closest to the classical state, it is instructive  to compare the result of
this formalism  with the classical gravity theory.

 Consider the oscillatory phase of  inflaton; the  initial values of
the inflaton
can be incorporated with the amplitude and the   phase of the real inflaton
\begin{equation}
\phi_{r}(t)=\frac{{\cal{B}}_{0}}{S^{3\over2}(t)\sqrt{w(t)}} \sin(\int w(t)dt+
\delta)\,\,,
\end{equation}
where ${\cal{B}}_0$ is the amplitude of the classical inflaton.

 The Hamiltonian for the classical inflaton is given by
\begin{equation}
H_{m}=\frac{1}{2S^{3}} \pi^{2}_{\phi_r} + \frac{m^{2}S^{3}}{2} \phi^{2}_{r}
\,\,,
\end{equation}

  $\pi^2_{\phi_r}$ and $\phi_{r}^2$  can be evaluated  and these values are
 substituted
in (59).   The classical energy density
for the inflaton in classical gravity becomes
\begin{eqnarray}
H_{m}&=&\frac{{\cal{B}}_{0}^{2}}{2} {1\over 2w}\left[m^{2}+w^{2}+{1\over4}\left
(\frac{\dot{w}}{w}+
3\frac{\dot{S}}{S}\right)^{2}+\left(w^{2}-{1\over4}\left(
\frac{\dot{w}}{w}+3\frac{\dot{S}}{S}\right)^{2}\right)\right.\nonumber\\
&&\times\cos(2\int w(t)dt+\delta)
-\left(\frac{\dot{w}}{w}
+3\frac{\dot{S}}{S}\right)\sin2(\int w(t) dt +\delta)\nonumber\\
&&-\left.m^{2}\cos2(\int w(t) dt)
\right]\,\,.
\end{eqnarray}
The result of the above Hamiltonian (60) shows the oscillatory
behavior of the classical energy density. The oscillating terms
 determine, in a significant way, the evolution of the geometric
invariants through the higher derivatives of the scale factor S(t).

 The time average over several oscillations of the Hamiltonian (60) can
be
computed by assuming that the expansion of the universe can be neglected during
each period of coherent oscillation and is obtained [7] as
\begin{eqnarray}
\langle H_{m} \rangle_{tavg} &=&\frac{{\cal{B}}_{0}^{2}}{2} {1\over 2w}\left[{1\over4}\left(\frac{\dot{w}}{w}+
3\frac{\dot{S}}{S}\right)^{2} + m^{2}
+w^{2}\right]\,\,.
\end{eqnarray}
Next goal is to compare the above classical  result with that of the
semiclassical energy  of the inflaton in the coherent state. Hence,
 consider the
Hamiltonian of the inflaton in coherent state in (29).
Substituting (31)
in (29),
 the energy density in the coherent state becomes:
\begin{eqnarray}
\langle\hat{H}_{m}\rangle_{cs}=&&\frac{1}{2w(t)}\left[(|\alpha|^{2}+{1\over2})
({\cal{A}}_{1}+w^{2}(t)+m^{2})\right. \nonumber\\
&&-{1\over2}\alpha^{*2}\exp(2i\int w(t)dt\left({\cal{A}}_{1}-w^{2}(t)-iw(t)
\left[3\frac{\dot{S}(t)}{S(t)}+\frac{\dot{w}(t)}{w(t)}\right]+m^{2}\right)
\nonumber\\
&&-\left.{1\over2}\alpha^{2}\exp(-2i\int w(t)dt\left({\cal{A}}_{1}-w^{2}(t)+iw(t)
\left[3\frac{\dot{S}(t)}{S(t)}+\frac{\dot{w}(t)}{w(t)}\right]
+m^{2}\right)\right]^{1\over3}\nonumber\\
\end{eqnarray}
where ${\cal{A}}_{1}$ is given by (47)\,\,.
 As in  the case of the classical energy density, the time average over
several oscillations of the energy density of the inflaton in coherent state
can be computed by assuming that the expansion of the universe can be neglected
during each period, and is obtained
as
\begin{eqnarray}
\langle \hat{H}_{m}\rangle_{cs-tavg}=&&\left(|\alpha|^{2}+{1\over2}\right)
\frac{1}{2w(t)}
\left[{1\over4}\left(\frac{\dot{w}}{w}+3\frac{\dot{S}}{S}
\right)^2+w^{2}(t)+m^{2}\right]\,\,.
\end{eqnarray}
 Comparing (61) with (63), we obtain the  classical amplitude:
\begin{equation}
{\cal{B}}_0=\sqrt{2|\alpha|^2+1}.
\end{equation}
{\section{Discussions and conclusions} }
 In this paper, we studied  particle production
of the coherently oscillating inflaton, after the inflation, in coherent
  and squeezed states formalisms, in the frame work of semiclassical theory
of gravity.
 The number of
particles at a later time $t$,  produced from the coherent state,
at the initial time $t_0$, in the limit $mt_0$, $mt > 1$
calculated. It show, the particle production depends on the
coherent  state parameter. The particle creation in squeezed state
in the limit   $mt_0 >mt >1$ is also computed , it is found that
the particle production depending on the associated squeezing
parameter. Similarly the number of particles produced in the
squeezed state  also  computed.
 It is observed
that, when $r=0$, the result matches with the number of particles produced in the
coherent state
 and when $\alpha = 0$, the result fits with the number of particles
produced in the  squeezed vacuum state.

    Quantum effects can play a significant role in the oscillatory phase
of the inflaton after inflation in the event that,  coherent and
squeezed states are  possible quantum states of the inflaton in that regime,
  the particle production can be affected the coherent oscillations of the
inflaton after the inflation. This
is became the fact that, the inflaton after inflation can not execute coherent
oscillation for a sufficiently long period of time, since it will suffer
from an instability due to the particle production. The particle production
in coherent state, squeezed vacuum state and squeezed state also depend on the duration of the coherent oscillations
of the inflaton.

In this paper, we have also studied  the inflaton in classical
gravity and compare the results with the inflaton in
semiclassical gravity by representing  it in coherent state
formalism, since the coherent states are  considered as the
closest to the classical states. From the comparative study of
the energy density of the inflaton in coherent state with the
classical energy density, it is found that the energy density in
coherent state also exhibit oscillatory behavior apart from  the
quantum correction. This could be due to the inherent quantum
nature of the coherent state. Hence, the semiclassical energy
density and classical energy density may  differ because of the
non-oscillatory nature of the  quantum correction of the energy
density in coherent state formalism.
 It can also be  concluded that, if the expansion
 of the
universe can
be neglected during each period of coherent  oscillation,
then the time-averaged classical
energy density,  in the leading order, is the same as that given by the
semiclassical gravity in the coherent state, if one  identifies the classical
amplitude with  ${\cal{B}}_0=\sqrt{2|\alpha|^2+1}$\,\,. A plot for coherent state
parameter verses  $ {\cal{B}}_0$ is presented in Fig.1.
\vskip0.5cm
\begin{center}
\input epsf
\centerline{
 \epsfxsize=4.0in   \epsfbox  {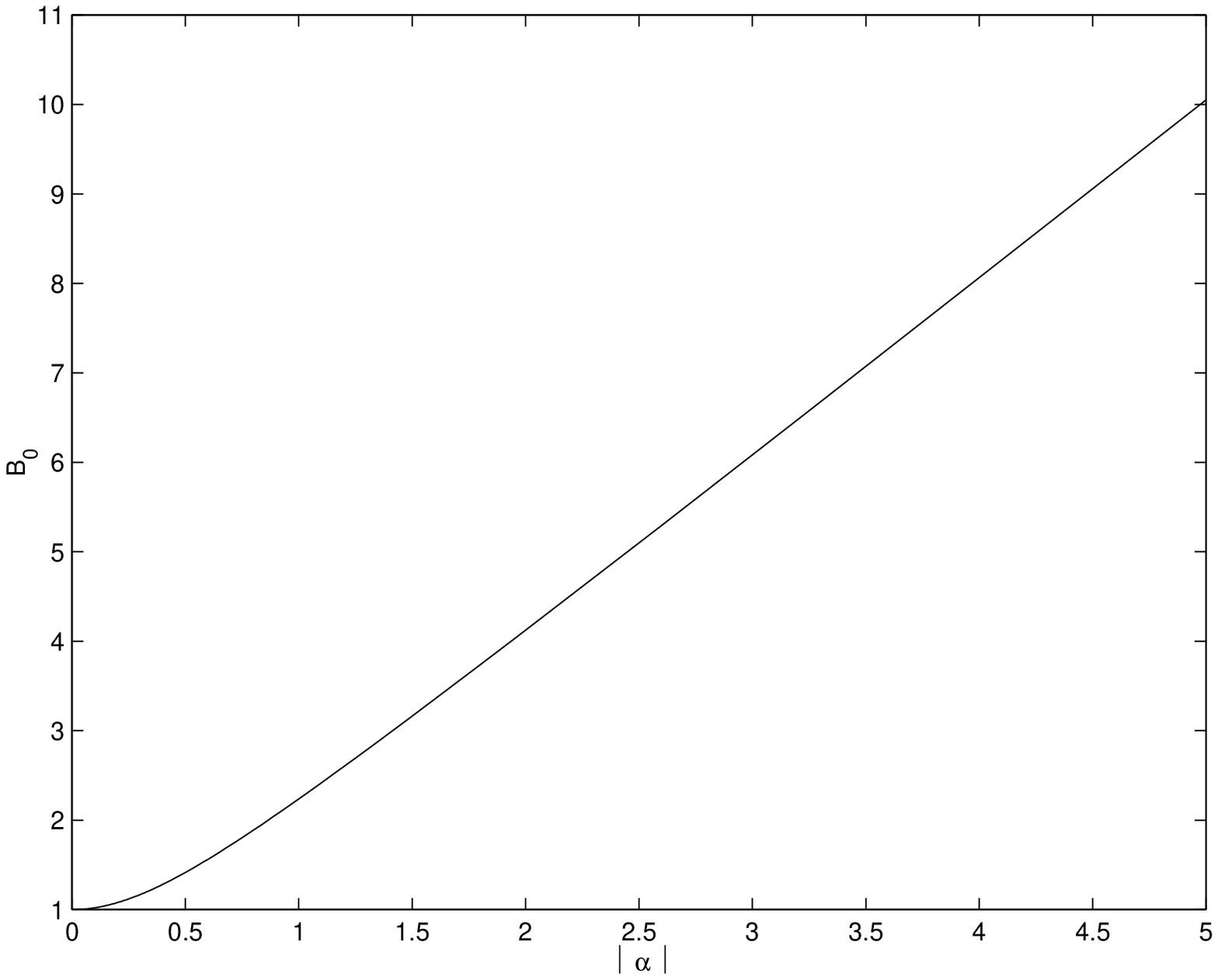} }
\vskip0.5cm
Fig.1.Plot for $| \alpha|$ with ${\cal{B}}_0$.
\end{center}
{\bf Acknowledgment}\\

P.K.S  wishes to thank the Director and IUCAA, Pune, India, for
warm hospitality and acknowledges Associateship of IUCAA.


\end{document}